\documentclass[letter]{aa}     % for the letters
\usepackage{graphicx}
\usepackage{placeins}
\usepackage{txfonts,url}
\usepackage{subcaption}         % necessary for continued figures, example in section 3
\usepackage{lscape,longtable}             % to rotate a single page table, example in appendix.
\usepackage{placeins}           % useful with \FloatBarrier, to keep 
\usepackage{booktabs}
\usepackage[table,xcdraw]{xcolor}
\usepackage[]{hyperref}
\newcommand{\kms}{km s$^{-1}$}

\begin{document}

%%%%%%%%%%%%%%%%%%%%%%%%%%%%%%%%%%%%%%%%
% if you use custom commands in your title,
% ensure to check your title when submitting!
%%%%%%%%%%%%%%%%%%%%%%%%%%%%%%%%%%%%%%%%
   \title{No country for old stars}

   \subtitle{Spectroscopic confirmation of the first intermediate-age RR Lyrae \\in the open cluster Trumpler 5}

%%%%%%%%%%%%%%%%%%%%%%%%%%%%%%%%%%%%%%%%
% Please separate each author with the \and command
%
% Please do not include ORCIDs next to author names.
% Only ORCIDs authenticated by individual authors in EDPS
% editorial system will be taken into account.
% ORCIDs included here will be removed.
%%%%%%%%%%%%%%%%%%%%%%%%%%%%%%%%%%%%%%%%

\author{V. D'Orazi\inst{1,2}\thanks{The first two authors contributed equally. Corresponding authors: vdorazi@roma2.infn.it; cmateu@fcien.edu.uy}%
        \and C. Mateu\inst{3}\protect\footnotemark[\value{footnote}]%
        \and G. Iorio\inst{4}
        \and A. Bobrick\inst{5,6}
        \and Z. Prudil\inst{7}
        \and R. Salinas\inst{8}
        \and A. Bragaglia\inst{9}
        \and L. Mashonkina\inst{10}
        \and R. G. Gratton\inst{11}
        \and I. Ilyn\inst{12}
        \and N. Alvarez Baena\inst{1,2}
        \and V. F. Braga\inst{2}
        \and A. Nunnari\inst{1,2}
        \and V. M. Kalari\inst{13}
        \and F. Cusano\inst{9}
        \and S. Tosi\inst{2}}        
   \institute{Department of Physics, University of Rome Tor Vergata, via della Ricerca Scientifica 1, I-00133, Rome, Italy
   \and
   INAF – Osservatorio Astronomico di Roma, Via Frascati 33, Monte Porzio Catone, I-00078, Italy.
   \and
   Departamento de Astronomía, Instituto de Física, Universidad de la República, Iguá 4225, CP 11400 Montevideo, Uruguay. 
   \and 
   Institut de Ci\`encies del Cosmos (ICCUB), Universitat de Barcelona, c. Mart\'{i} i Franqu\`{e}s 1, 08028 Barcelona, Spain
   \and
   School of Physics and Astronomy, Monash University, Clayton, VIC 3800, Australia
   \and
    Australian Research Council Centre of Excellence for Gravitational Wave Discovery, Clayton, VIC 3800, Australia
    \and
    European Southern Observatory, Karl-Schwarzschild-Strasse 2, D-85748 Garching bei M\"unchen, Germany
    \and
    Nicolaus Copernicus Astronomical Center, Polish Academy of Sciences, Bartycka 18, 00-716 Warszawa, Poland
    \and
    INAF -  Osservatorio di Astrofisica e Scienza dello Spazio di Bologna, via P. Gobetti 93/3, I-40129 Bologna, Italy 
    \and
    Institute of Astronomy, Russian Academy of Sciences, 119017 Moscow, Russia
    \and
    INAF - Osservatorio Astronomico di Padova, vicolo dell'Osservatorio 5, I-35122, Padova, Italy
    \and
    Leibniz-Institut f\"ur Astrophysik Potsdam, An der Sternwarte 16, D-14482 Potsdam, Germany
    \and
    Gemini Observatory/NSF’s NOIRLab, Casilla 603, La Serena, Chile}   

   \date{Received: April 6 2026. Accepted: April 10 2026}

% \abstract{}{}{}{}{}
% 5 {} token are mandatory
 
  \abstract
  % conteTelescopext heading (optional)
   {RR Lyrae (RRL) stars are widely considered tracers of ancient ($>$10 Gyr) metal-poor stellar populations. However, recent kinematic and photometric studies suggest the existence of a metal-rich RRL subpopulation associated with the thin disk and intermediate ages ($\sim$2–5 Gyr), therefore challenging canonical evolutionary models.} 
   {We aim to provide the first spectroscopic confirmation of a member of this elusive population. Specifically, we target a metal-rich RRL candidate recently identified photometrically as a member of the intermediate-age open cluster Trumpler 5 ($\sim$2.5 Gyr).} 
   {We obtained high-resolution spectroscopy using PEPSI at the LBT and GHOST at Gemini South telescope. We measured radial velocities (RVs) from multiple epochs to constrain cluster membership and derived detailed chemical abundances (Mg, Ca, Sc, Ti, Mn, Fe, Cu, Zn, Y, and Ba) to compare the RRL's composition with that of red clump stars in the cluster.}
   {The RRL's systemic velocity ($V_\gamma = 50.57^{+0.78}_{-0.36}$ km s$^{-1}$) is in excellent agreement with the cluster mean ($V = 50.76\pm0.49$ km s$^{-1}$).  Combining RVs, proper motions, and parallax, the probability of the star being a background interloper is negligible ($\sim0.002\%$, better than $4\,\sigma$). We derived a metallicity of [Fe/H] = $-0.40 \pm 0.05$, which matches the cluster value. While most abundance ratios (Mg, Ti, Mn, Cu, and Zn) align with cluster members, the RRL exhibits significant depletion in Ca, Sc, Y, and Ba. Notably, [Sc/Fe] is underabundant by $\sim$0.6 dex relative to the cluster stars, following trends seen in field metal-rich RRLs.}
   {We provide strong constraints on the membership status between an RRL variable and an intermediate-age open cluster. Cluster membership enables accurate measurement of RRL age and chemical anomalies relative to its host, particularly in Sc and neutron-capture elements. These anomalies further reinforce a nonstandard formation channel for this RRL, possibly indicating binary interactions and mass transfer.
   }


   \maketitle
   \nolinenumbers

%%%%%%%%%%%%%%%%%%%%%%%%%%%%%%%%%%%%%%%%%%%%%%%%%%%%%%%%%%%%%%
\section{Introduction}
RR Lyrae stars (RRLs) are low‑mass core–helium‑burning pulsators located on the horizontal branch of the Hertzsprung–Russell diagram \citep{preston1964}. Occupying the classical instability strip, they exhibit radial pulsations with characteristic periods on the order a few tenths of a day (typically $<$ 1 day; e.g., \citealt{catelan2015}) and provide important tests of stellar evolution and pulsation theory. In addition, RRLs are well-established standard candles. Their utility as distance indicators is underpinned by tight period–luminosity relations \citep[see, e.g.,][]{narloch2024}. This property makes them essential for accurate distance determinations in the Milky Way and nearby galaxies as well as for tracing the formation and evolutionary history of the Galactic components \citep[e.g.,][]{neeley2019, bono2026}. Canonical single-star evolutionary models suggest that RRLs are produced only in stellar systems that are both very old (ages $\gtrsim$ 9–10 Gyr) and sufficiently metal poor ([Fe/H] $\lesssim -0.5$). Under these conditions, the horizontal branch extends across the instability strip, allowing stars to enter the pulsation regime characteristic of RR Lyrae variables (e.g., \citealt{catelan2015}). The high incidence of RRLs in predominantly ancient environments—such as Galactic globular clusters, ultra-faint dwarf galaxies, and stellar halos—has reinforced the view that they trace exclusively old, metal-poor stellar populations. Beginning with \cite{Layden1995} and further supported by \cite{zinn2020}, \cite{prudil2020}, and \cite{Iorio2021}, a substantial subset of metal-rich RRLs ([Fe/H] $\gtrsim -0.5$, extending to near-solar metallicities) has been identified. These stars exhibit kinematics consistent with a relatively young thin-disk component characterized by markedly colder and more rapidly rotating Galactic orbits than the $>10$ Gyr disk populations expected for canonical RRLs. Additional support comes from recent analyses of RRL kinematics in the warped outer disk \citep{Cabrera-Gadea2025} and from comparisons between the kinematics of RR Lyrae and Mira variables of different ages, with the latter inferred from the Mira period–age relation \citep{Zhang2025}.  Furthermore, \cite{cuevas2025} identified 23 probable RR Lyrae members in ten intermediate-age (1–8 Gyr) clusters in the Large and Small Magellanic Clouds. However, the authors emphasize that spectroscopic confirmation is still required to secure their cluster membership. 

Despite this converging evidence, current claims for intermediate-age RRLs remain based on indirect age diagnostics and population-statistical arguments rather than direct age determinations. Very recently, \cite{2025arXiv250922336M} (hereafter M25) reported the first detection of an RRL (Gaia DR3 3326852328563919744) associated with the intermediate‑age (2–4 Gyr) open cluster (OC) Trumpler 5, found by cross‑matching a large all‑sky RRL compilation with kinematically selected members of over 3000 Gaia‑based OCs. The multicolor light curves, parallax, proper motions, position within the cluster tidal radius, and location in the Gaia color–magnitude diagram are all consistent with cluster membership, with only a $\sim 0.05\%$ (or $3.5\,\sigma$) probability of being a field interloper (see M25 for details). Nonetheless, the lack of spectroscopic constraints on its kinematics and chemical composition has so far prevented a fully secure association. In this paper, we present high-resolution spectroscopic observations of Mateu’s star and derive its radial velocity (RV) and detailed chemical abundances, conclusively confirming its membership in Trumpler 5.
%
%%%%%%%%%%%%%%%%%%%%%%%%%%%%%%%%%%%%%%%%%%%%%%%%%%%%%%%%%%%%%%

%%%%%%%%%%%%%%%%%%%%%%%%%%%%%%%%%%%%%%%%%%%%%%%%%%%%%%%%%%%%%%
\section{Observations and analysis}
The target RRL star was observed with the high-resolution PEPSI spectrograph \citep{strassmeier2015} on the Large Binocular Telescope (Mount Graham, Arizona) between October 2025 and February 2026 as part of a DDT program.  
The PEPSI observations used simultaneous spectral setups CD1+CD5 or CD2+CD4 (see Table \ref{tab:data}), yielding R $\approx 50000$; all spectra were used to measure RVs. However, CD1 and CD2 suffered from very low S/N, and the remaining CD4 or CD5 setups contain a limited number of usable lines because the RRL spectral energy distribution peaks toward blue wavelengths ($\sim 400$ nm). 
To obtain chemical abundances, we therefore relied on additional observations with the new high-resolution GHOST spectrograph \citep{Kalari2024} at Gemini South (Cerro Pachón, Chile). GHOST delivers R $\approx$ 45000 (2×2 binning) and continuous coverage from 347–1061 nm with its blue and red arms (see Table \ref{tab:data}). Exposures were limited to 1200 s; for parameter and abundance determination, we used only the spectrum obtained at  $ \phi \approx$ 0.3 —the optimal phase for RRL abundance analysis (e.g., \citealt{For2011})— which also has the highest S/N (highlighted in yellow in Table \ref{tab:data}). The determination of the RVs for each spectrum was performed using the cross-correlation algorithm in the 2023 version of \texttt{iSpec} \citep{blanco2014}. 

Stellar parameters and elemental abundances were determined using \texttt{PySME} \citep{pysme2023}, a Python implementation of Spectroscopy Made Easy \citep{Piskunov2017}. This software allows for real-time computation of  Non Local Thermodynamical Equilibrium (NLTE) synthetic spectra and employs a least-squares minimization procedure using the dogbox algorithm as implemented in \texttt{scipy} \citep{Virtanen2020}. The code provides error estimates for each parameter and abundance by calculating random and systematic uncertainties. The internal random error, which is dependent on the S/N of the spectra, can be readily determined from the least-squares fit in PySME using the covariance matrix. 
Systematic uncertainties were computed following the methodology outlined in \cite{Piskunov2017}, but they are not relevant for this study, as our primary objective is to perform a differential analysis of the composition of our RRL relative to red clump (RC) stars in Trumpler 5. For our analysis, we employed model atmospheres from the MARCS grid \citep{gustafsson2008} and NLTE departure coefficients from \cite{Amarsi2020} for all elements except Ti and Cu (which were sourced from \citealt{Mallinson2024} and \citealt{caliskan2025}, respectively) and Sc. The departures from LTE for Sc were specifically calculated for this study by following the approach described in \cite{Mashonkina2022}. Abundances for Zn and Y are the only ones we report in LTE since departure coefficient grids are not available within the PySME framework. 

We optimized the effective temperature ($T_{\rm eff}$), surface gravity ($\log g$), microturbulence ($V_{\rm mic}$), and macroturbulence ($V_{\rm mac}$), along with the abundances of iron and titanium, using a comprehensive line list that includes Fe {\sc i}, Fe {\sc ii}, Ti {\sc i}, and Ti  {\sc ii} lines (see Table \ref{tab:linelist} for details). An example of our spectral synthesis, for a restricted range in wavelengths, is shown in Figure \ref{fig:spectral_synthesis}. Subsequently, we adopted the derived stellar parameters and metallicity to infer the elemental abundances of Mg, S, Ca, Sc, Ti, Mn, Cu, Zn, Y, and Ba. 
To minimize the impact of systematic uncertainties in comparing abundances of our target RRL and the OC members, we retrieved two RC stars from the ESO archive, initially published in \cite{Donati2015}. These spectra were obtained with UVES \citep{Dodorico2000} at the Very Large Telescope (VLT), featuring a nominal resolution of R$\simeq 40,000$ (slit width of 1.2 arcsec) and a spectral coverage of $\lambda\lambda$ 328–456 nm (CD\#2) and 472–683 nm (CD\#3). We analyzed these two stars using exactly the same methodology as for the RRL (i.e., identical line lists, model atmospheres, and radiative transfer code), thereby reducing systematic effects. The only exception is sulfur. The high-excitation triplet at 6743–6757 \AA~ is too weak to be detected in the RC spectra, preventing an estimate of their S abundance.
%

%%%%%%%%%%%%%%%%%%%%%%%%%%%%%%%%%%%%%%%%%%%%%%%%%%%%%%%%%%%%%%
\section{Results and discussion}

\paragraph{Radial velocities -} The star’s systemic line-of-sight velocity ($V_\gamma$) was obtained by fitting the RV template from \citet{Prudil2024} to the phase-folded observations. The amplitude of the RV curve ($\mathrm{AmpRV}$) and an additional velocity dispersion term ($\sigma_x$) were included as free parameters. Full details about the inference model can be found in Appendix~\ref{a:rv_curve}, including the phase-folded RV curve and residuals with respect to the best-fitting model. The best-fitting values found were $V_\gamma = 50.57_{-0.36}^{+0.78}$~\kms, $\mathrm{AmpRV} = 54.6_{-3.2}^{+3.1}$~\kms, and $\sigma_x=1.82_{-0.35}^{+0.45}$~\kms, which correspond to the maximum a posteriori (MAP) and uncertainties to 14th and 86th percentiles, respectively.

The systemic RV found for the Trumpler 5 RRL is in excellent agreement with the cluster mean reported by \citet{Ozdemir2025}:  50.76~\kms\,, with a velocity dispersion of $\sigma_v=0.49~$\kms. We can now use this result to compute an updated probability that a background RRL would be misidentified as a cluster member. The authors of M25 used RRLs within 25\degr of the cluster to model their distribution in parallax and proper motion and obtained a probability of 0.0049\% that a random background star would have the same parallax and proper motion as the Trumpler 5 RRL. We included an RV factor, estimated from the background distribution from the selection of plausible RRL stars ($M_{bol}\in[0.45,0.7]$ and $T_\mathrm{eff}\in[6200,7800]$~K) from the Gaia Universe Model and Gaia Object Generator simulations, and we modeled it with a Gaussian decomposition. This yielded a probability of $p=0.0004$\% that a random background RRL will have the same parallax, proper motion, and RV as ours (to within three times its uncertainties and the cluster's velocity dispersion). Finally, considering there is one possible member out of a total of four RRLs within the cluster’s tidal radius, we obtained a binomial probability of 0.0017\% for the star to be a random background interloper.

\paragraph{Elemental abundances -} The stellar parameters and elemental abundances for our RRL, along with RC stars in Trumpler 5, are detailed in Table~\ref{tab:abu_rrl}. 
For the two RC stars, our retrieved stellar parameters and [Fe/H] align closely with those of \cite{Donati2015}, reinforcing the notion that significant systematic errors do not adversely affect our results. With the exception of manganese and barium—the former exhibits a deficit of approximately $\approx -$0.2 dex, while the latter shows an enhancement of about 0.25 dex—all other species display solar-scaled [X/Fe] ratios, consistent with expectations for Population~{\sc i} stars. The enhancement of Ba in intermediate-age and young open clusters has been a persistent issue in the field, as discussed in \citet[][and references therein]{dorazi22}. We note, in passing, that the supersolar [Ba/Fe] values supports our adopted age of $\tau=2.5$ Gyr rather than the older value of around 4 Gyr (see the extensive discussion in M25).

\begin{figure*}
    \centering
    \includegraphics[width=0.85\linewidth]{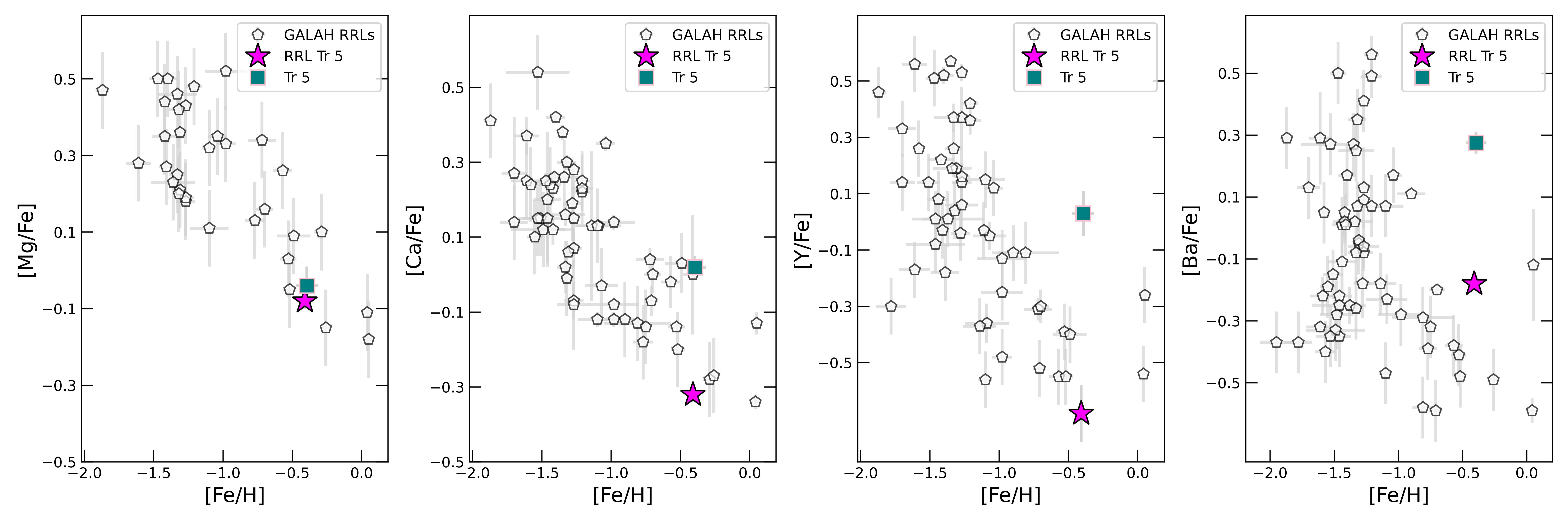}
    \caption{[X/Fe] for Mg, Ca, Y, and Ba as a function of [Fe/H] for the RRL, the average of two RC stars in Trumpler 5, and for RRLs in the GALAH sample by \cite{dorazi2024}.} 
    \label{fig:XFe_galah}
\end{figure*}

For the RRL star, we derived a metallicity of [Fe/H] = $-0.40 \pm 0.05$, in excellent agreement with the cluster value. When comparing abundance ratios, we found very good agreement in Mg, Ti, Mn, Cu, and Zn between the RRL and the OC stars (see Figure \ref{fig:comparison_all}). In contrast, Ca, Sc, Y, and Ba show significant discrepancies: The RRL exhibits notably subsolar [X/Fe] ratios for these elements. Such differences in Ca, Y, and Ba between metal-rich RRLs and field giants and dwarfs have already been reported in the literature, including the recent GALAH DR3 sample of RRLs analyzed by \cite{dorazi2024}. However, that work (as all the previous ones) compared field RRLs with field nonvariable stars, mixing populations of different ages, metallicities, and other abundances. Our study is the first to compare the chemical peculiarities of a metal-rich RRL directly with stars from the same cluster, thereby using well-constrained parent abundances and removing the mixed-population-related systematics. Figure~\ref{fig:XFe_galah} shows the comparison of [X/Fe] for Mg, Ca, Y, and Ba of the target with RRLs in GALAH DR3. Mateu’s star lies in a region consistent with field RRLs at a similar metallicity that exhibit subsolar [X/Fe] in Mg, Ca, Y, and Ba. Moreover, we found a striking underabundance in Sc: [Sc/Fe] = $-0.40 \pm 0.05$ for the RRL versus [Sc/Fe] = $+0.18 \pm 0.05$ for the OC stars. 
Scandium was not analyzed in the previous work by \cite{dorazi2024}. However, \cite{chadid2017} reported a strong declining trend of [Sc/Fe] with [Fe/H] in their sample of RRLs, with metal-rich RRLs showing a significant Sc underabundance (see  Figure~\ref{fig:scandium_chadid}). Our star follows this trend closely, matching their results. 

The origin of these unusual abundances remains unclear, but we speculate that they may be linked to the formation scenario of young metal-rich RRLs. In particular, binary interactions and mass transfer involving a circumbinary disk could produce chemical peculiarities similar to those observed in post-AGB stars \citep{kamath2019,mohorian2025}. Making further progress on this topic will require quantitative modeling of circumbinary disks (including their structure, evolution, and dust–gas separation processes) coupled with binary-evolution calculations tailored to young metal-rich RRL progenitors. Even more important will be the analysis of statistically significant samples of RR Lyrae stars in order to robustly characterize the incidence and abundance patterns of the observed anomalies. These aspects will be addressed in a forthcoming study.

\paragraph{Formation scenario -} 
Our study offers definitive confirmation, verified through kinematics and elemental abundances, of the association between the RRL identified by M25 and the intermediate-age cluster Trumpler 5 ($\tau =2.5$ Gyr). This association poses significant challenges in understanding the formation of these objects at such a young age, as single stellar evolution models are insufficient. To explain the properties of metal-rich RR Lyrae stars in the Galactic bulge and in relatively metal-rich globular clusters, He enrichment is often proposed (see, e.g., \citealt{savino2020}). However, confirmation of He enrichment based on pulsation properties has not been obtained in any of these environments (\citealt{marconiminniti2018,li2025}). In addition, He enrichment models generally predict higher luminosities and longer periods compared to field RRLs (see, e.g., \citealt{marconi2018,bhardwaj2022}). Indeed, all RRLs in metal-rich globular clusters exhibit longer periods than field RRLs of similar metallicity \citep{pritz2000,Cruz2024}. This is not consistent with Mateu's star. 

\cite{Bobrick2024} predicted the formation of metal-rich RR Lyrae stars as a consequence of binary interactions (see also \citealt{karczmarek2017}). Key predictions from these models are that most metal-rich RR Lyrae stars, particularly the younger cohort (ages below 9 Gyr), currently reside in low-mass binary systems with orbital periods ranging from approximately 900 to 2000 days. While this is consistent with the Galactic kinematics and formation rates of most metal-rich RR Lyrae stars (see, e.g., \citealt{Zhang2025,Cabrera-Gadea2025}), and while the same model also agrees with the detailed observations of binary hot subdwarfs \citep{Vos20, Molina26}, a definitive confirmation of RRL binary companions is still lacking, including in Gaia DR3 data (\citealt{iorio2026}; see also \citealt{pranav2026}). For the current mass of M$_{\text{rrl}}$ = 0.55 M$_{\odot}$, the potential companion mass of 1.5 M$_{\odot}$, and a period of P=2000 days suggested by the model assuming the cluster age, the semi-amplitude of the RV variation (K) of RRL is 15.71 km s$^{-1}$ (for a circular orbit with an inclination of 90 degrees). Given our approximately 120-day observational time span, we expect a signal of around 2.9 km s$^{-1}$. This signal magnitude remains consistent with the additional scatter of roughly $\sim$ 2 km s$^{-1}$ required to fit the RV template, although it allows us to somewhat exclude the presence of more massive companions and/or shorter orbital periods. We plan to conduct a rigorous RV follow-up in the forthcoming months to better constrain possible binary properties. 

\begin{acknowledgements}
Based on observations collected with PEPSI at the Large Binocular Telescope (DDT observations on October 2025 - February 2026; GHOST at Gemini (November 2025-February 2026, programs GS-2025B-Q-415 and GS-2026A-Q-116); UVES at VLT (program 074.D-0344).
Our thanks to Chris Sneden and M\'arcio Catelan for insightful comments and suggestions. Based on observations obtained at the international Gemini Observatory, a program of NSF NOIRLab, which is managed by the Association of Universities for Research in Astronomy (AURA) under a cooperative agreement with the U.S. National Science Foundation on behalf of the Gemini Observatory partnership: the U.S. National Science Foundation (United States), National Research Council (Canada), Agencia Nacional de Investigaci\'{o}n y Desarrollo (Chile), Ministerio de Ciencia, Tecnolog\'{i}a e Innovaci\'{o}n (Argentina), Minist\'{e}rio da Ci\^{e}ncia, Tecnologia, Inova\c{c}\~{o}es e Comunica\c{c}\~{o}es (Brazil), and Korea Astronomy and Space Science Institute (Republic of Korea).This work made extensive use of CDS and NASA ADS databases. AB, VD acknowledge support from  INAF Minigrant 2022, 2024. CM acknowledges funding from RDT and CSIC MIA programs from the Universidad de la República, Uruguay. GI was supported by a fellowship grant from the la Caixa Foundation (ID 100010434). AB acknowledges support from the Australian Research Council (ARC) Centre of Excellence for Gravitational Wave Discovery (OzGrav), through project number CE230100016. The fellowship code is LCF/BQ/PI24/12040020. AN acknowledges the contribution of NextGenerationEU funds within the National Recovery and Resilience Plan (PNRR), Mission 4-Education and Research, Component 2-From Research to Business (M4C2), Investment Line 3.1-Strengthening and creation of Research Infrastructures, Project IR0000034-"STILES", CUP C33C22000640006. ST acknowledges support from the INAF research project LBT–Supporto Arizona Italia. 
\end{acknowledgements}
\bibliographystyle{aa} % style aa.bst
\bibliography{tr5_rrl.bib} % your references Yourfile.bib
\begin{appendix}
\nolinenumbers 

%\onecolumn
\section{Log of observations and radial velocities}
In Table \ref{tab:data} we report information on the acquired spectra with PEPSI and GHOST, including the Julian Date, the phase of the spectra, the wavelength range, the average S/N ratios per pixel and the RV values.
\begin{table*}[htbp]
    \centering
    \caption{Spectroscopic observations alongside pulsational phase ($\Phi$) and individual RV measurements with associated uncertainties. The two spectra analyzed for abundances are highlighted.}\label{tab:data}
    \small
    \begin{tabular}{lccccr}
        \hline
        Spectrum ID &  Julian & $\Phi$ & Wavelength & S/N & RV \\
                &       Date &       &     range [nm]   & [\text{pix}$^{-1}$] & [km s$^{-1}$] \\
        \hline
                &  \hspace{2cm} & \textbf{PEPSI at LBT} & & & \\
        pepsib.20251025.000 & 2460973.935966 & 0.340 &  382-426 & 10  & $48.61\pm0.14$ \\
        pepsir.20251025.000 & 2460973.935966 & 0.340 &  624-742 & 110 & $48.85\pm0.15$ \\
        pepsib.20251025.001 & 2460973.957165 & 0.380 &  422-477 & 20  & $52.53\pm0.18$ \\
        pepsir.20251025.001 & 2460973.957165 & 0.380 &  536-627 & 103 & $52.60\pm0.16$ \\
        pepsib.20251214.056 & 2461023.782983 & 0.327 &  383-426 & 8 & $47.20\pm0.80$ \\
        pepsir.20251214.036 & 2461023.782983 & 0.327 &  624-742 & 106 & $49.16\pm0.78$ \\
        pepsib.20251214.057 & 2461023.804165 & 0.368 &  422-477 & 20  & $51.79\pm0.25$ \\
        pepsir.20251214.037 & 2461023.804165 & 0.368 &  536-627 & 113 & $51.31\pm0.25$ \\
        pepsir.20251214.040 & 2461023.894412 & 0.540 &  536-627 & 15 & $60.14\pm0.28$ \\
        pepsir.20260205.002 & 2461076.780448 & 0.310 & 536-627 & 50 & 
        $47.31\pm0.48$ \\
        pepsir.20260205.003 & 2461076.811965 & 0.370 & 536-627 & 50 & 
        $51.99\pm0.48$ \\
            &  \hspace{2cm} & \textbf{GHOST at Gemini-S} & & & \\
        S20251109S0153\_blue & 2460988.846577  & 0.754 & ~347-544 & 20 & $74.58\pm0.34$ \\
        S20251109S0153\_red  & 2460988.846577  & 0.754 & 521-1061 & 40 & $74.97\pm0.45$ \\
        S20251111S0090\_blue & 2460990.835672  & 0.544 & ~347-544 & 14 & $62.46\pm 0.26$ \\
        S20251111S0090\_red  & 2460990.835672  & 0.544 & 521-1061 & 45 & $62.84\pm0.37$ \\
\rowcolor{yellow!40} S20260211S0052\_blue & 2461082.539757 & 0.284 & ~~347-544 & 80 & $44.54\pm0.20$ \\
\rowcolor{yellow!40} S20260211S0052\_red & 2461082.539757 & 0.284 & 521-1061 & 110 & $44.87\pm0.20$ \\
        \hline
    \end{tabular}
\end{table*}

\section{Line list and spectral synthesis}
\begin{figure}[htbp]
    \centering
    \includegraphics[width=0.95\linewidth]{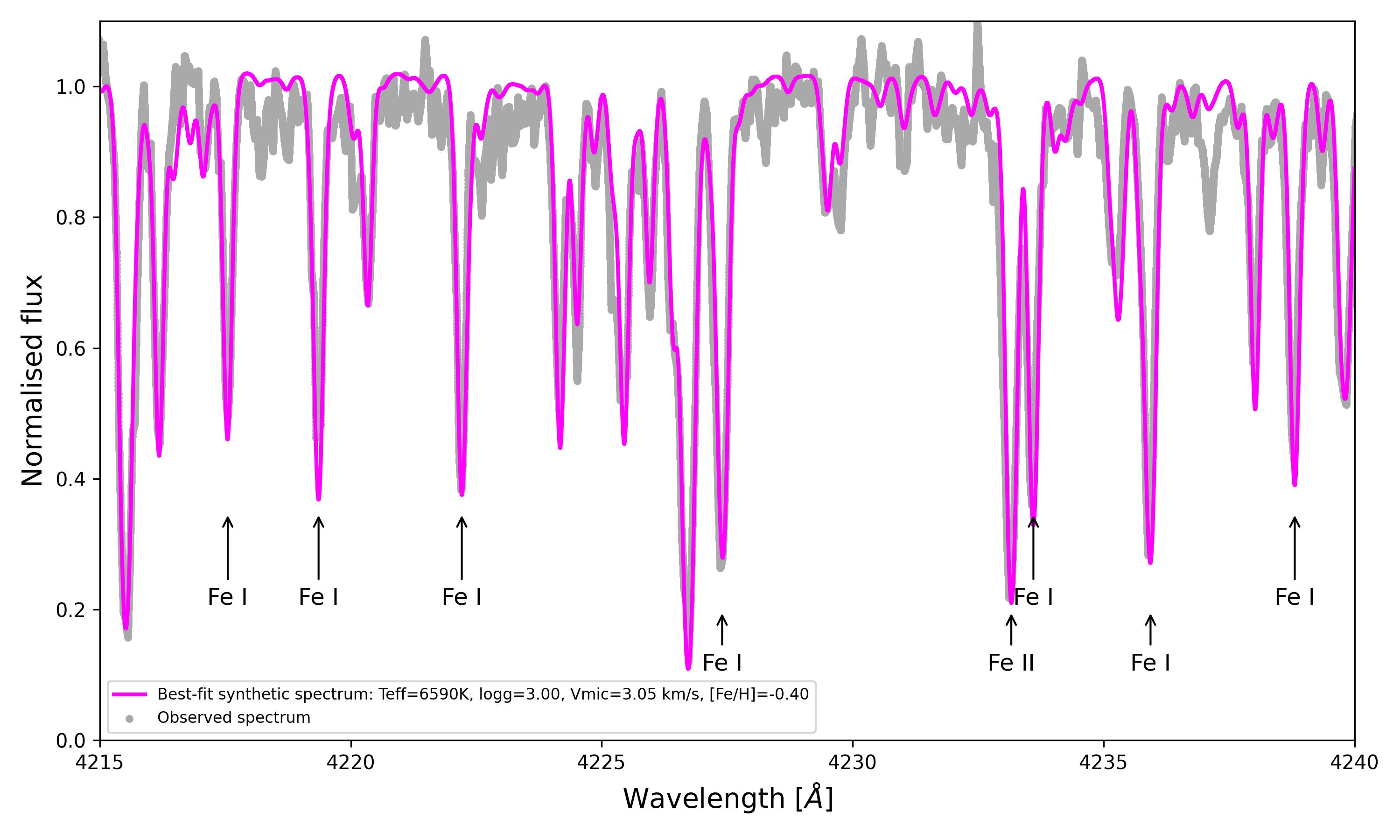}
    \caption{Comparison between observed (gray points) and synthetic (magenta continuous line) spectrum.}
    \label{fig:spectral_synthesis}
\end{figure}

For the derivation of atmospheric parameters and stellar abundances, the line list provided in Table \ref{tab:linelist} was utilized. This table contains the wavelengths, species, excitation potentials (E.P.), and transition probabilities ($\log gf$) in columns 1, 2, 3, and 4, respectively. In Figure \ref{fig:spectral_synthesis} we show an example of the spectral fitting for the GHOST spectrum of our RRL. The figure compares the observed GHOST spectrum with the best-fit synthetic spectrum computed with PySME, obtained by simultaneously fitting the stellar atmospheric parameters and [Fe/H]. Several Fe I and Fe II lines within this narrow wavelength interval are marked.

\begin{table}[htbp]
\caption{Line list used for atmospheric parameter and elemental abundance determinations.}
\label{tab:linelist}
\begin{tabular}{lccr}
\hline
Wavelength & Ion & E.P. & $\log gf$ \\
\text{[\AA]} &         & \text{[eV]} & \\
& & & \\
4217.545 & Fe 1 & 3.43 & -0.48 \\  
4219.359 & Fe 1 & 3.57 & 0.00 \\  
4222.213 & Fe 1 & 2.45 & -0.97 \\  
4240.784 & Fe 1 & 4.37 & -2.35 \\  
4271.760 & Fe 1 & 1.48 & -0.16 \\  
4282.403 & Fe 1 & 2.18 & -0.78 \\  
4352.734 & Fe 1 & 2.22 & -1.29 \\  
4375.929 & Fe 1 & 0.00 & -3.03 \\  
4383.544 & Fe 1 & 1.48 &  0.20 \\  
4388.407 & Fe 1 & 3.60 & -0.68 \\  
4404.750 & Fe 1 & 1.56 & -0.14 \\  
4427.309 & Fe 1 & 0.05 & -2.92 \\  
4432.567 & Fe 1 & 3.57 & -1.60 \\  
4447.717 & Fe 1 & 2.22 & -1.34 \\  
4514.184 & Fe 1 & 3.05 & -2.05 \\  
4574.215 & Fe 1 & 3.21 & -2.50 \\  
4602.001 & Fe 1 & 1.61 & -3.15 \\  
4602.940 & Fe 1 & 1.48 & -2.21 \\  
4607.645 & Fe 1 & 3.98 & -3.24 \\  
4619.288 & Fe 1 & 3.60 & -1.12 \\  
4630.120 & Fe 1 & 2.28 & -2.59 \\  
4647.434 & Fe 1 & 2.95 & -1.35 \\  
4661.970 & Fe 1 & 2.99 & -2.50 \\  
5217.389 & Fe 1 & 3.21 & -1.07 \\  
5364.870 & Fe 1 & 4.45 & 0.23 \\  
5569.618 & Fe 1 & 3.42 & -0.49 \\  
5576.089 & Fe 1 & 3.43 & -1.00 \\  
5638.262 & Fe 1 & 4.22 & -0.87 \\  
5679.023 & Fe 1 & 4.65 & -0.90 \\  
5705.464 & Fe 1 & 4.30 & -1.36 \\  
5778.453 & Fe 1 & 2.59 & -3.43 \\  
5855.075 & Fe 1 & 4.61 & -1.48 \\  
5883.816 & Fe 1 & 3.96 & -1.36 \\  
5916.247 & Fe 1 & 2.45 & -2.99 \\  
5956.693 & Fe 1 & 0.86 & -4.61 \\  
6056.004 & Fe 1 & 4.73 & -0.46 \\  
6151.617 & Fe 1 & 2.18 & -3.30 \\  
6165.359 & Fe 1 & 4.14 & -1.47 \\  
6240.646 & Fe 1 & 2.22 & -3.23 \\  
6246.318 & Fe 1 & 3.60 & -0.73 \\  
6315.811 & Fe 1 & 4.08 & -1.71 \\  
6336.823 & Fe 1 & 3.69 & -0.86 \\  
6380.743 & Fe 1 & 4.19 & -1.38 \\  
6494.980 & Fe 1 & 2.40 & -1.27 \\  
\end{tabular}
\end{table}

\begin{table}[htbp]
\begin{tabular}{lccr}
\hline
Wavelength & Ion & E.P. & $\log gf$ \\
\text{[\AA]} &         & \text{[eV]} & \\
& & & \\
4233.162 & Fe 2 & 2.58 & -1.90 \\  
4273.320 & Fe 2 & 2.70 & -3.30 \\  
4296.566 & Fe 2 & 2.70 & -2.93 \\  
4385.377 & Fe 2 & 2.78 & -2.68 \\  
4416.819 & Fe 2 & 2.78 & -2.41 \\  
4491.397 & Fe 2 & 2.86 & -2.70 \\  
4508.280 & Fe 2 & 2.86 & -2.25 \\  
4515.333 & Fe 2 & 2.84 & -2.45 \\  
4520.218 & Fe 2 & 2.81 & -2.60 \\  
4541.516 & Fe 2 & 2.86 & -2.79 \\  
4576.333 & Fe 2 & 2.84 & -2.92 \\  
4582.830 & Fe 2 & 2.84 & -3.09 \\  
4583.829 & Fe 2 & 2.81 & -1.86 \\  
4620.513 & Fe 2 & 2.83 & -3.24 \\  
4629.331 & Fe 2 & 2.81 & -2.33 \\  
4993.350 & Fe 2 & 2.81 & -3.64 \\  
5264.802 & Fe 2 & 3.23 & -3.12 \\  
6084.102 & Fe 2 & 3.20 & -3.78 \\  
6149.246 & Fe 2 & 3.89 & -2.72 \\  
6432.676 & Fe 2 & 2.89 & -3.52 \\  
4913.613 & Ti 1 & 1.87 & 0.22 \\  
4981.731 & Ti 1 & 0.85 & 0.57 \\  
5219.702 & Ti 1 & 0.02 & -2.22 \\  
5689.460 & Ti 1 & 2.30 & -0.36 \\  
5978.541 & Ti 1 & 1.87 & -0.31 \\  
6091.171 & Ti 1 & 2.27 & -0.45 \\  
4874.009 & Ti 2 & 3.09 & -0.86 \\  
5211.530 & Ti 2 & 2.59 & -1.41 \\  
5381.022 & Ti 2 & 1.57 & -1.97 \\  
5418.768 & Ti 2 & 1.58 & -2.13 \\  
5528.405 & Mg 1 & 4.35 & -0.50 \\  
5711.088 & Mg 1 & 4.35 & -1.72 \\  
6743.580 & S 1 & 7.87 & -1.03 \\  
6748.790 & S 1 & 7.87 & -0.53 \\  
6757.150 & S 1 & 7.87 & -0.24 \\  
5588.749 & Ca 1 & 2.53 & 0.36 \\  
6122.217 & Ca 1 & 1.89 & -0.32 \\  
6161.297 & Ca 1 & 2.52 & -1.27 \\  
5526.770 & Sc 2 & 1.77 & 0.02 \\  
5640.989 & Sc 2 & 1.50 & -1.13 \\  
5657.906 & Sc 2 & 1.51 & -0.54 \\  
6013.475 & Mn 1 & 3.07 & -0.25 \\  
6016.643 & Mn 1 & 3.07 & -0.20 \\  
6021.747 & Mn 1 & 3.08 & 0.06 \\  
5105.542 & Cu 1 & 1.39 & -1.54\\
4722.153 & Zn 1 & 4.03 & -0.34 \\  
4883.682 & Y 2 & 1.08  & -0.30 \\  
5853.669 & Ba 2 & 0.60 & -3.01 \\  
6141.709 & Ba 2 & 0.70 & -2.26 \\
\end{tabular}
\end{table}

\section{Comparison with cluster abundances}
\begin{figure}[h]
    \centering
\includegraphics[width=0.5\textwidth]{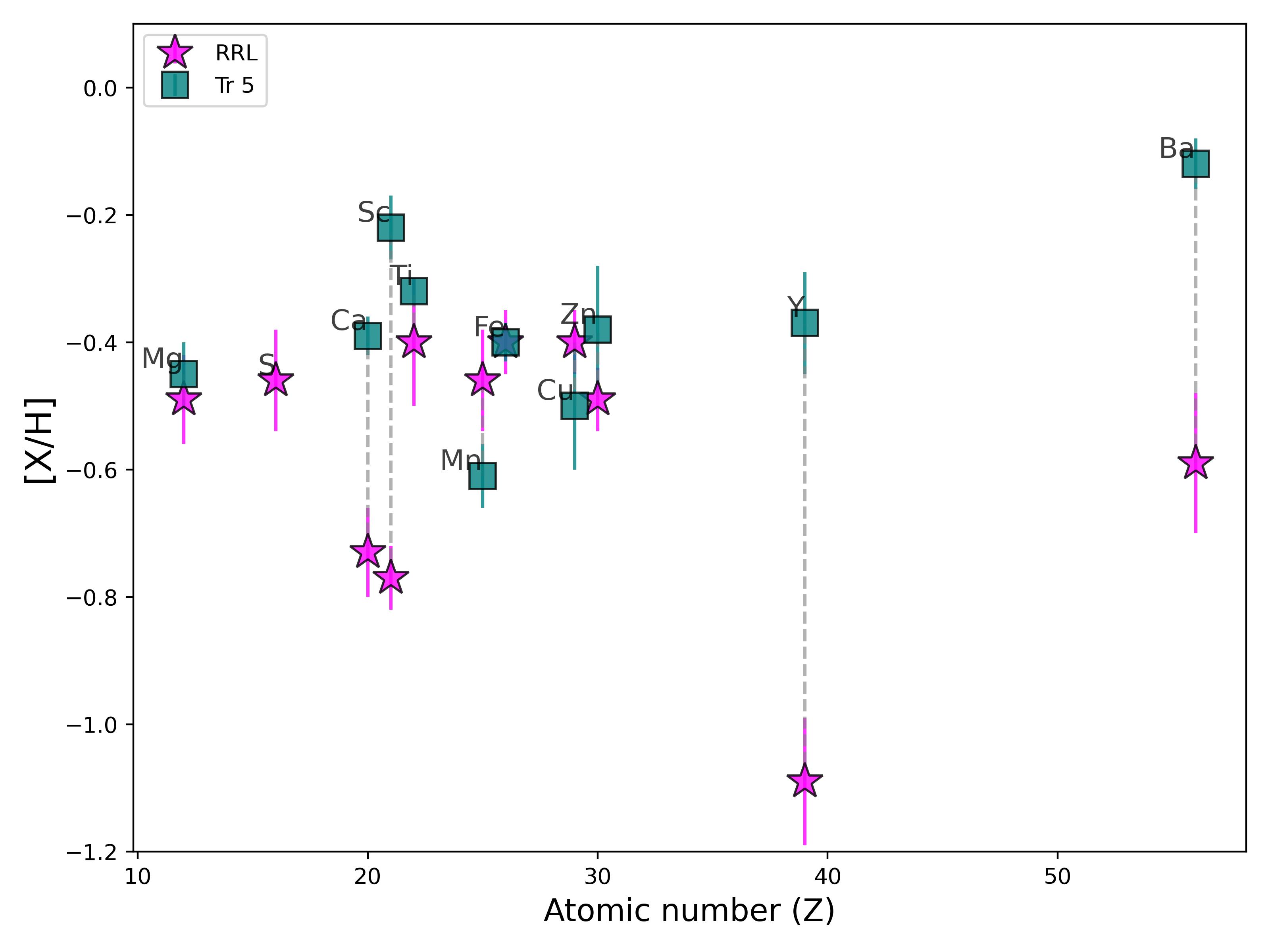}
    \caption{Comparison between the RRL and open cluster Trumpler 5 for the full suite of inferred elemental abundances.}
    \label{fig:comparison_all}
\end{figure}

The atmospheric parameters and elemental abundances for our sample of RRL stars and the two RC stars in Trumpler 5 are detailed in Table \ref{tab:abu_rrl}. Columns 1,2, and 3 are for the RRL Gaia DR3 3326852328563919744 and two RC stars in Trumpler 5, respectively. All elements are in NLTE except for Zn and Y since departure coefficients are not available in PySME. The errors represent internal uncertainties associated with the best-fit procedures. The errors on the mean values have been conservatively estimated as the average of the individual errors.
Figure \ref{fig:comparison_all} compares the [X/H] abundance ratios of Mateu's star with the OC mean derived from the two RC stars. While the overall abundance patterns are consistent, there are distinct discrepancies in Ca, Sc, Y, and Ba. We attribute these differences to the peculiar origin of this star.

\begin{table}[htbp]
\centering
\small
\caption{Stellar parameters and abundance ratios ([X/H]).  }\label{tab:abu_rrl}
{\small \begin{tabular}{l|ccc}
\hline
 &  RRL &  {\small Cl* Tr5 }  & {\small Cl* Tr 5} \\ 
 &  {\small GHOST@Gemini}    &  {\small K 3-2553}  &   {\small K 3-3236}\\
 \hline
T$_{\rm eff}$  [K]  & 6590$\pm$150   & 4793$\pm$90  & 4863$\pm$60  \\
$\log g$         & 3.00$\pm$0.10  & 2.39$\pm$0.08 & 2.71$\pm$0.10    \\
$V_{\rm mic}$ [km s$^{-1}$]  & 3.50$\pm$0.20  & 1.33$\pm$0.20 & 1.40$\pm$0.20   \\ 
$V_{\rm mac}$ [km s$^{-1}$]  & 9.50$\pm$0.50  & 4.53$\pm$0.26 & 4.89$\pm$0.38    \\
\text{[Mg/H]} & $-0.49\pm0.02$ & $-0.45\pm0.10$ & $-0.42\pm0.10$   \\
\text{[S/H]}  & $-0.46\pm0.08$ &                   &                 \\
\text{[Ca/H]} & $-0.73\pm0.02$&   $-0.40\pm0.04$ & $-0.37\pm0.06$   \\
\text{[Sc/H]} & $-0.80\pm0.08$&   $-0.25\pm0.05$  & $-0.18\pm0.04$    \\
\text{[Ti/H]} & $-0.40\pm0.03$&   $-0.35\pm0.08$  & $-0.29\pm0.06$   \\
\text{[Mn/H]} & $-0.46\pm0.05$&   $-0.61\pm0.10$  & $-0.60\pm0.08$  \\
\text{[Fe/H]}       & $-0.40\pm0.05$ &  $-0.42\pm0.10$ & $-0.37\pm0.06$   \\
\text{[Cu/H]} & $-0.40\pm0.05$ & $-0.48\pm0.08$ & $-0.51\pm0.09$  
\\
\text{[Zn/H]} & $-0.49\pm0.05$ & $-0.46\pm0.10$ & $-0.30\pm0.10$  \\
\text{[Y/H]} & $-1.09\pm0.10$ & $-0.47\pm0.11$ & $-0.26\pm0.09$  \\
\text{[Ba/H]} & $-0.59\pm0.10$&   $-0.15\pm0.08$  &  $-0.09\pm0.06$   \\
\hline
\end{tabular}
}
\end{table}
\section{Scandium abundances}
\begin{figure}[h!]
    \centering
    \includegraphics[width=0.95\linewidth]{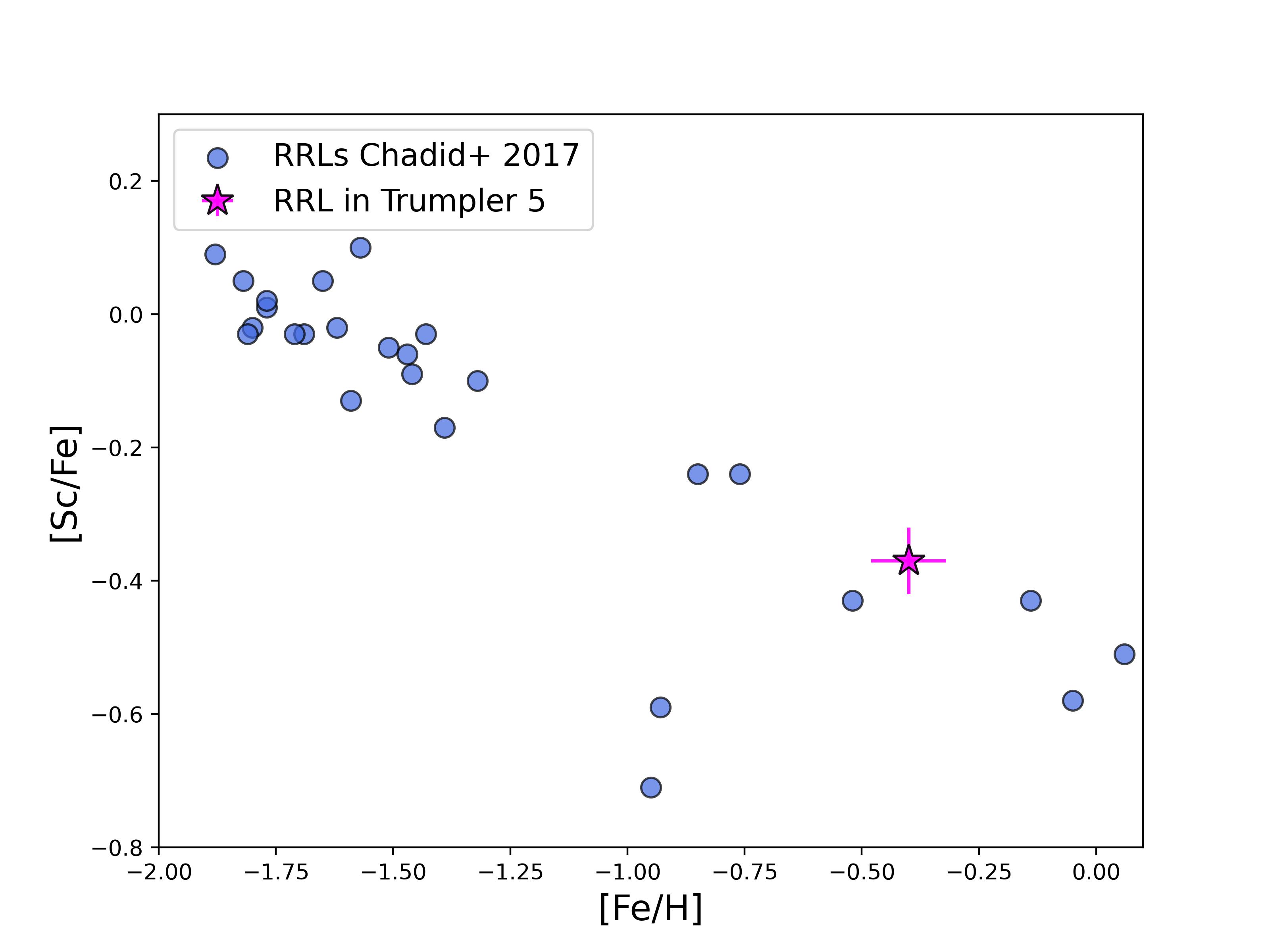}   
    \caption{[Sc/Fe] as a function of metallicity for the RRL sample by \cite{chadid2017} and our RRL in Trumpler 5.}
    \label{fig:scandium_chadid}
\end{figure}

In Figure \ref{fig:scandium_chadid}, we present the [Sc/Fe] ratio of our star alongside the RRab sample from \cite{chadid2017}. 
We observe that our target RRL, notable for its significant 
subsolar Sc content, aligns well with the abundance trend defined by field metal-rich RRLs. The NLTE corrections, omitted in \cite{chadid2017}, are negligible for Sc II lines at the atmospheric parameters (temperature, gravity, metallicity) characteristic of metal-rich RRLs ($\lesssim$ 0.05 dex).

\section{Systemic velocities}\label{a:rv_curve}

The RV data was phase-folded assuming the star's period $P=0.524807$~d and epoch at maximum light (HJD$_0$=2457000.44368~d) reported by the OGLE-IV database, and transformed to Barycentric Julian Date. The OGLE-IV data were chosen as the reference since it provides the best sampled light curve currently available for the star (see Fig.~1 in M25).  
 
The RV template from \citet{Prudil2024} was fitted to the phase-folded RV data, assuming a simple model with Gaussian uncertainties. Following \citet{Prudil2024}, we take the zero-point of the fit as the systemic velocity of the star, corresponding to the point of zero RV in the template occurring at $\phi=0.35$. We also treat the amplitude of the RV curve, $\mathrm{AmpRV}$, as a free parameter in our model, taking advantage of the good phase coverage provided by our observations. We include an extra velocity dispersion $\sigma_{x}$ term (added to the RV uncertainties in quadrature) to account for a possible underestimation of the RV errors and/or additional (possibly physical) sources of uncertainty. Uniform priors were assumed for all parameters, in the ranges $V_\gamma\in[-500,500]$~\kms, $\mathrm{AmpRV}\in[0,150]$~\kms and $\sigma_x\in[0,50]$~\kms. 

\begin{figure*}[htbp]
    \centering
    \includegraphics[width=0.45\linewidth]{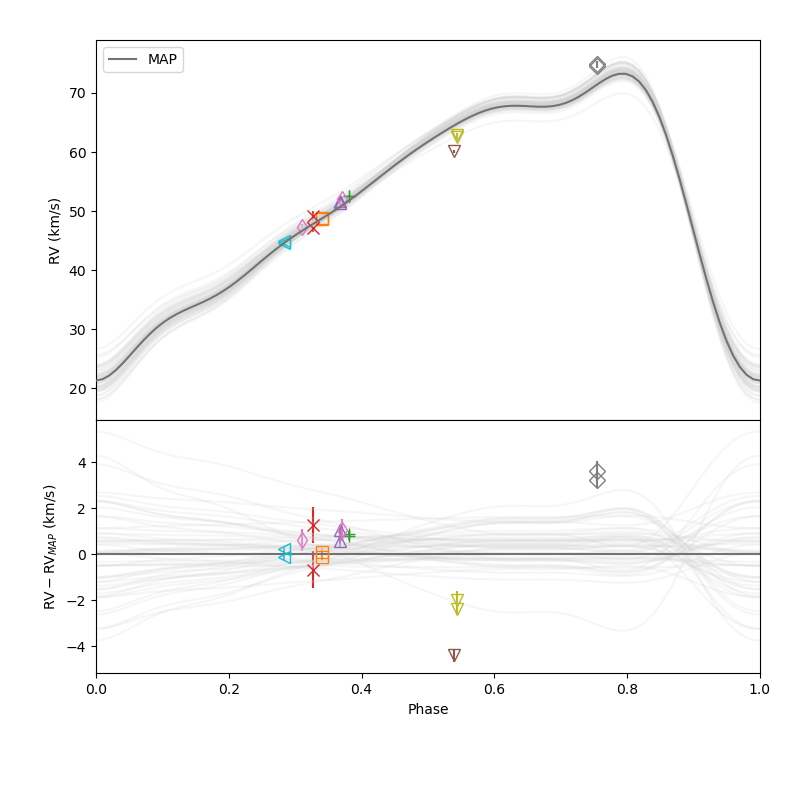}  
    \includegraphics[width=0.45\linewidth]{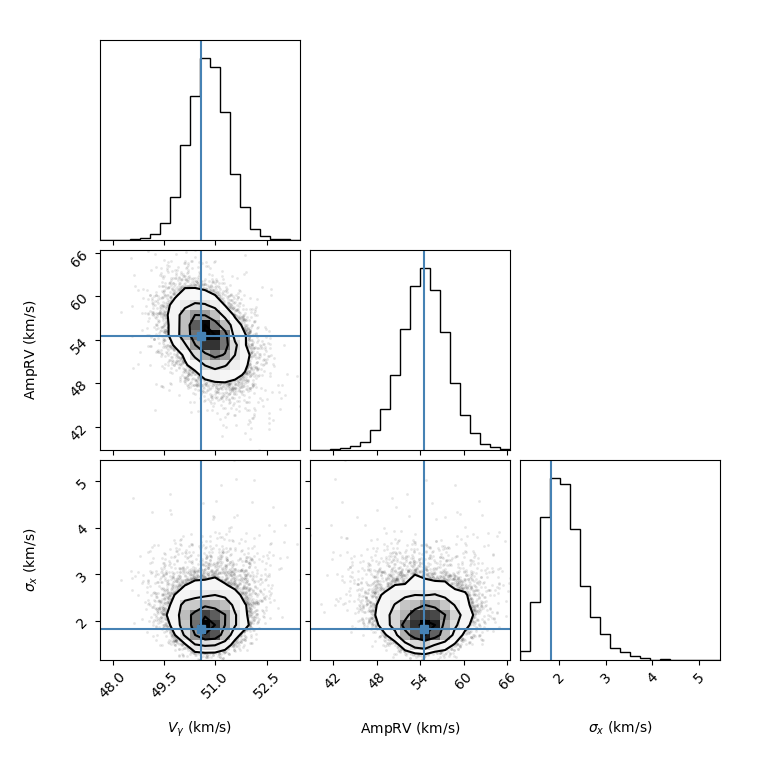} 
\caption{Radial velocity (top) and best-fit residuals (bottom) as a function of phase. The best-fitting template (MAP) is shown by the dark solid line. Light gray lines correspond to 50 random draws from the posterior PDF. Different symbols correspond to different observation epochs, with the two observations corresponding to the spectra from the blue/red channels at the same epoch sharing the same symbol. Error bars for the RVs are shown, but are typically smaller than the symbols. \emph{Right panel:} Corner plot showing the marginal 2-D and 1-D posterior PDFs for the three free parameters: systemic velocity $V_\gamma$, amplitude of the RV curve AmpRV and the additional velocity dispersion term $\sigma_x$. The solid (blue) line corresponds to the MAP of the (joint) 3D posterior PDF.}
\label{fig:rv_fit}
\end{figure*}

Figure~\ref{fig:rv_fit} shows the phase-folded RV curve with the best-fitting model (left plot, top panel) and residuals (left plot, bottom panel) and the corner plot with 1-D and 2-D posterior probability density functions (PDF) for the fit parameters (right plot). The best-fitting value is taken as the maximum-a-posteriori (MAP) of the 3D (joint) PDF, and the uncertainties reported correspond to the 14th and 86th percentiles of the marginal distributions, respectively.

\section{Color-magnitude diagram}

\begin{figure}[h]
    \centering
    \includegraphics[width=0.9\linewidth]{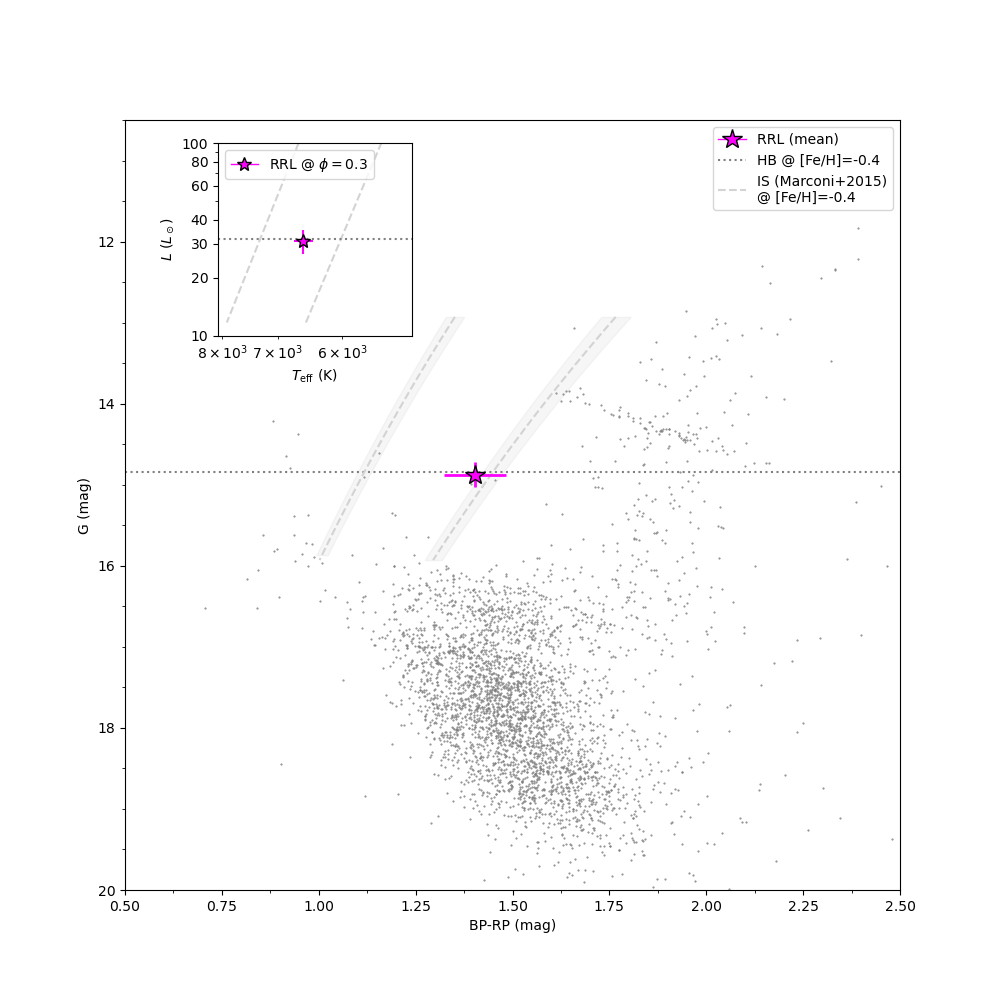}    
\caption{Color-magnitude diagram for the Trumpler~5 OC. The mean position (intensity-averaged magnitude and color) of the RRL is shown with the (magenta) star. Cluster members from \citet{CantatGaudin2020} are shown as (gray) dots. The dashed and dotted lines correspond to the instability strip limits from \citet{marconi2015} and horizontal branch magnitude from the \citet{Garofalo2022} period-luminosity-metallicity relation at the star's metallicity, distance and extinction of $A_V=1.65\pm 0.15$ following M25 (see their Sec.3 for details). The insert shows the H-R diagram and the position of the star at $\phi=0.3$ }
    \label{fig:cmd}
\end{figure}

Figure~\ref{fig:cmd} shows the color-magnitude diagram in the Gaia bands for the cluster population. The insert shows the H-R diagram in a zoomed-in view around the RRL star at $\phi\approx0.3$, the phase at which the star's temperature was inferred. This shows the RRL is well inside the instability strip and supports the adoption of the extinction ($A_V=1.65$) from the \citet{Green2019} 3D dustmap from M25, rather than the lower extinction ($A_V=1.5$) predicted from the \citet{lallement2022} dust map which placed the star just past the edge of the instability strip.  

\end{appendix}

\end{document}